\documentclass[10pt,osajnl2,letterpaper,twocolumn,showpacs]{revtex4}  %% REVTeX 4.0

\usepackage{graphicx} % standard LaTeX graphics tool
\usepackage{amsmath}   % AMS tex
\usepackage{amssymb}   % AMS tex symbols
\usepackage{color}

\begin{document}

\newcommand {\edit}[1]{\textcolor{red}{#1}}

\newcommand {\note}[1]{\textcolor{blue}{\textbf{#1}}}

\title{Atom trapping in an interferometrically generated  bottle beam  trap}

\author{L. Isenhower, W. Williams, A. Dally, and M. Saffman}
\affiliation{Department of Physics, University of Wisconsin, 1150 University Avenue,  Madison, Wi., 53706}

\begin{abstract}
We demonstrate an optical bottle beam  trap created by interfering two fundamental Gaussian beams with different waists. The 
beams are derived from a single laser source using a Mach-Zehnder interferometer whose arms have unequal magnifications.  Destructive interference of the two beams from the Mach-Zehnder leads to a three dimensional intensity null at the mutual focus of the beams. We demonstrate trapping of cold Cesium atoms 
in a blue detuned  bottle beam trap. 
\end{abstract}

%\date{\today}


\maketitle %% null function with osajnl.sty

The use of neutral atoms for quantum information processing is a topic of great current interest. 
The quantum states of single atoms can be manipulated by first localizing them in tightly confining  traps. A simple method to achieve this is a far off-resonance optical  trap (FORT). This trap can take two forms: a bright trap using red detuned light where the atoms are attracted to a region of high intensity\cite{Miller1993}, or a dark trap using blue detuned light based on repulsion of the atoms from a surrounding region of high  intensity. 
The dark trap has several advantages over the bright trap. The atoms (or other trapped particles) are held in an area of low intensity and therefore scatter fewer photons when compared to a bright trap of the same depth. This significantly decreases the atomic heating and decoherence rates. In addition trap induced AC Stark shifts are minimized which simplifies coherent control of internal states using additional optical fields. We are especially interested in dark traps for application to coherent control of Rydberg atoms\cite{Johnson2008} since the wavelength can be chosen to give equal trapping potentials for both 
ground and Rydberg states\cite{Saffman2005}.

Several methods have been used to produce bottle beams that have an intensity null surrounded by light in all directions\cite{Ozeri1999,ref.BoB1,ref.capped,ref.axicon,Yelin2004,ref.toroid}. Some of these approaches require optical access from several sides, or the use of custom optical polarization plates, holograms, or spatial light modulators. Here we demonstrate a method that
is closest to that of Ref. \cite{Yelin2004} 
 to produce a bottle beam using a single laser source with a Gaussian TEM$_{00}$ spatial profile, and a Mach-Zehnder interferometer. Our approach
 can readily handle powers of several Watts and can be  scaled from $\mu\rm m$ to cm sized traps simply by changing the final focusing lens.

%%%%%%%%%%%%%%%%%%%%%%%%%%%%%%%%%%%%%%%%%%%%%%%%%%%%%
\begin{figure}[!t]
\includegraphics[width=8.1cm]{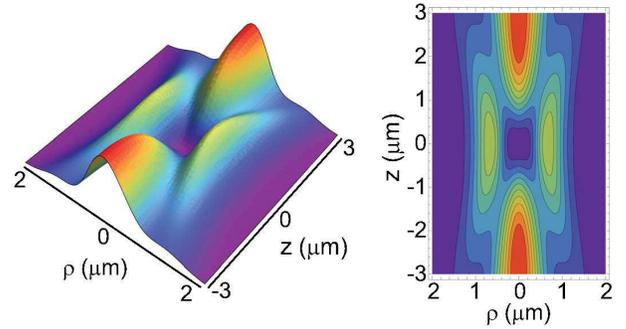}
\vspace{-.4cm}
\caption{(color online) Bottle beam intensity distribution for  $w_1=0.5~\mu\rm m,$ $w_2=0.94~\mu\rm m,$ and $\lambda=0.532~\mu\rm m$.  }
\label{fig.intensity_plot}
\end{figure}

We create a  bottle beam (BoB) trap by destructive interference of two lowest order Gaussian beams with different waists.
The beams of wavelength $\lambda$ propagating along $\hat z$ are assumed to have the same polarization and are both focused with their waists $w_1, w_2$ located at $z=0.$ Provided the beam powers satisfy $P_1 /w_1^2=P_2/w_2^2$ there is an intensity null at $x=y=z=0.$ Using standard expressions for TEM$_{00}$ Gaussian beams 
we find the axially symmetric intensity pattern shown in Fig. \ref{fig.intensity_plot}. Surrounding the 
intensity zero there are axial maxima at $z_m=\pm \pi q w_1^2 /\lambda$ and a radial maximum at $\rho_m=\sqrt2  q w_1 \left( \ln(q)/(q^2-1)\right)^{1/2},$ where we have introduced $q=w_2/w_1$. The BoB provides a three dimensional trapping potential, with the lowest escape barrier at the saddle points on the line $\theta=\tan^{-1}(z/\rho)\sim 20~\rm deg.$ As we show below the 3D localization is optimized for 
$q=q_0\simeq 1.89.$ Setting $q=q_0$, as in Fig. \ref{fig.intensity_plot}, the peak axial intensity is 
$I_{\rm max}\simeq 0.069 \frac{2P}{\pi w_1^2}$ with $P=P_1+P_2$ the total power. 
The radial intensity
and the   intensity at the saddle point reach, respectively, 61\% and 32\% of $I_{\rm max}$.

The spatial localization of cold atoms in the BoB trap  is an important figure of merit for quantum information applications.   
To quantify the localization we evaluate the optical intensity near the null along the axial
and radial   directions as 
\begin{eqnarray}
I(\rho=0,z)&=& \frac{2P \lambda^2}{\pi^3 w_1^6}\frac{(1-q^2)^2}{(1+q^2)q^4}z^2 + {\mathcal O}(z^4)\\
I(\rho,z=0)&=& \frac{2P }{\pi w_1^6}\frac{(1-q^2)^2}{(1+q^2)q^4}\rho^4 + {\mathcal O}(\rho^6).
\end{eqnarray}
Let us assume $w_2>w_1$ so $q>1$, then for given values of the power, wavelength, and smaller waist $w_1$ the localization in both 
$z$ and $r$ is strongest when $q=q_0=\sqrt{(3+\sqrt{17})/2}\simeq 1.89.$ 
The dipole potential for an atom with scalar polarizability $\alpha$ is $U=-\frac{\alpha}{2\epsilon_0 c}I$, where $c$ is the speed of light.  
The axial motion is harmonic and the oscillation frequency for an atom of mass $m$ and polarizability $\alpha$ is given by $\omega_z=\left(2\alpha \tilde q \lambda^2 P/(\pi^3 c \epsilon_0 w_1^6 m)\right)^{1/2}$ where $m$ is the atomic mass and we have introduced $\tilde q= (1-q^2)^2/[(1+q^2)q^4].$  For an atom with kinetic temperature 
$T$ the root mean square axial displacement is $z_{\rm rms}=\sqrt{k_B T/m\omega_z^2}$ with $k_B$ the Boltzmann constant.  

In the radial direction the potential is quartic and we can use the virial theorem to write 
$\langle \rho^4\rangle=\frac{2}{3U_{\rho4}}k_B T$ where $U_{\rho4}$ is the coefficient of $\rho^4$ in the expansion of the potential. 
Assuming a Maxwell-Boltzmann distribution of atoms in the quartic potential we find 
$\langle\rho^2\rangle/\langle \rho^4\rangle=\frac{2}{\sqrt\pi}
\left(\frac{U_{\rho 4}}{k_B T}\right)^{1/2}$
and $\rho_{\rm rms}=\frac{2}{3^{1/2}\pi^{1/4}} (k_B T/U_{\rho4})^{1/4}.$
Using the parameters of Fig. \ref{fig.intensity_plot}, $\alpha\times \frac{10^6}{4\pi\epsilon_0}=32\times 10^{-24}~\rm cm^3 $ for Cs atoms at $\lambda=0.532~\mu\rm m$, $T=10~\mu\rm K$, and  $P=10~\rm mW$
we find $z_{\rm rms}=0.28~\mu\rm m$ and $\rho_{\rm rms}=0.22~\mu\rm m$.  We can compare this with a red detuned FORT created by focusing a single beam to a waist $w_1$. Using $\lambda=1.064~\mu\rm m$, $w_1=1~\mu\rm m$, $\alpha\times \frac{10^6}{4\pi\epsilon_0}=170\times 10^{-24}~\rm cm^3 $  and the same temperature and power as for the BoB trap we find\cite{Saffman2005} $z_{\rm rms}=0.16~\mu\rm m$ and $\rho_{\rm rms}=0.055~\mu\rm m$. The bottle beam gives somewhat worse localization than the red detuned FORT, largely because the polarizability is much smaller at the shorter wavelength. Nevertheless we can readily obtain spatial localization to much better than one $\mu\rm m$ in all three dimensions with a power of only 10 mW and a much lower photon scattering rate than in the FORT. It is also worth noting that the motional decoherence properties of the BoB trap are slightly better than the FORT
due to the quartic nature of the radial potential.

\begin{figure}[!t]
\includegraphics[width=8.2cm]{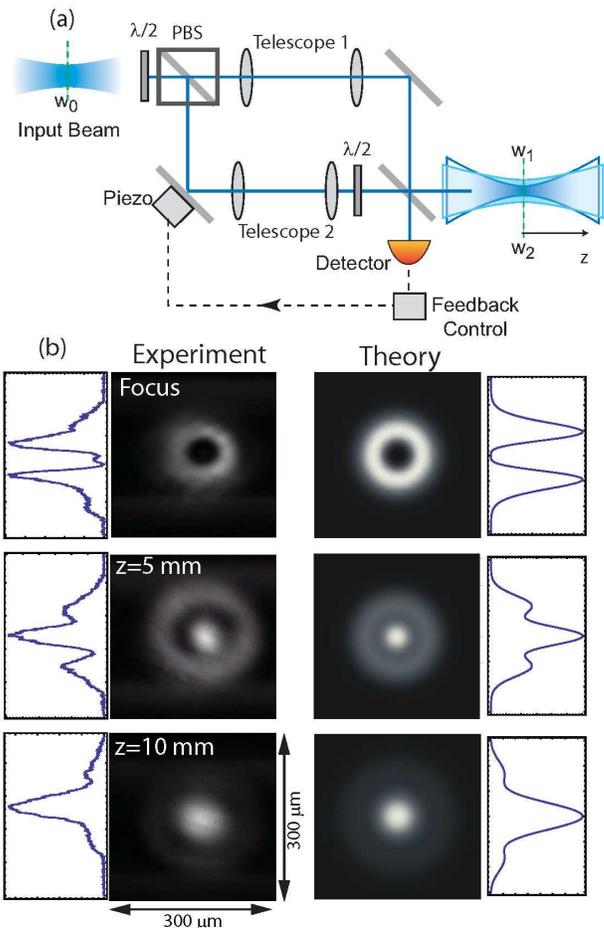}
\vspace{-.4cm}
\caption{(color online) (a)Schematic drawing of the interferometer used to generate the two beams. (b)Cross-sections of the intensity along the beam path for beam waists of  $w_1=37~\mu\rm m$ and $w_2=63~\mu\rm m$.}
\label{fig.center}
\end{figure}

We have demonstrated trapping of cold Cs atoms in a BoB trap using the  approach shown in Fig. \ref{fig.center}a.  A Mach-Zehnder interferometer  generates two beams with different waists by using telescopes of different magnification in each arm.  A feedback circuit is used to lock the two beams out of phase by controlling a piezoelectric mirror mounted in one arm. The BoB created at $z=0$ can then be reimaged into 
the desired experimental location with relay lenses.  This setup allows variation of the trap aspect ratio by varying the ratio of the two beam waists, or variation of the trap size by changing the magnification of 
the relay optics. 

To demonstrate the feasibility of the method we used an injection locked Ti:Sapphire laser to produce light detuned 40 GHz to the blue of  the $|6S_{1/2},F=4\rangle \rightarrow|6P_{3/2}\rangle$ D2 transition of Cs. The light was coupled to the Mach-Zehnder interferometer using a single mode polarization preserving fiber. Each arm of the interferometer contained a telescope to change the beam size. These telescopes had a magnification of .75x and 1.33x  giving a waist ratio of $q=1.78$ which is not far from the ideal $q_0$ discussed above. One output port of the interferometer was used to lock the two beams out of phase by maximizing the signal hitting a photodiode monitoring that output. To lock to this maximum a small dither was applied to the piezo controlled mirror for use with a lock-in amplifier. 
%The signal from the photodiode was sent to the lock-in and the output from the lock-in was used as an error %signal which was passed through a filter circuit which then controlled the piezo mounted mirror in order to %maintain the maximum. 
The other output went through a telescope (2.5x magnification) and a final focusing lens ($f=150~\rm  mm$) to focus the two beams onto a Magneto Optical Trap (MOT).  We obtained approximately 40 mW of trapping light at the atoms split between the two beams. The beam waists were $w_{1,2}=12, 22~\mu\rm m$ ($w$ is defined as the beam radius at the $1/e^2$ intensity point). 
The power was split between the two beams by adjusting a $\lambda/2$ plate before the input to the interferometer to obtain an intensity null at the focus.

We took images of the bottlebeam to measure the output waists and to compare them with the theoretical predictions. To do this we used a lens and CCD camera system, which had a magnification of 11, to image various axial planes of the bottlebeam. First we looked at each beam individually to measure each beam's waist and waist position. This allowed us to verify that each beam was focused at the same plane. The measured values for the beam waists were then used to create theoretical plots for comparison with pictures of both beams together. The results are shown in Fig. \ref{fig.center}b.

\begin{table}[!t]
\centering
\begin{tabular}{|l|c|}
\hline
 Quantity & BoB Trap\\
\hline
wavelength & $\lambda=.852 ~\mu\rm m$\\
optical power & $P=40~\rm  mW$\\
detuning & $\Delta/2\pi = +40~\rm  GHz$\\
beam waists & $w_{1,2}=12,22~\mu\rm m$\\
trap depth (axial,radial,saddle) & $4.4,2.7,1.4~\rm  mK$\\
position of axial intensity peak &$z_m=\pm 970 ~\mu\rm m$\\
position of radial intensity peak &$\rho_m=16 ~\mu\rm m$\\
spatial confinement  & $(z_{\rm rms}, \rho_{\rm rms})=(31.,3.0)~\mu\rm m$\\
axial vibrational frequency & $\omega_z/2\pi = 290~\rm  Hz$\\
%Radial Vibrational Frequency & 3500 Hz\\
photon scattering rate & $730~\rm s^{-1}$\\
\hline
\end{tabular}
\vspace{-.35cm}
\caption{Trap parameters calculated from our measured beam waists and power for 50 $\mu\rm K$ Cs atoms.}
\label{tab.parms}
\end{table}

\begin{figure}[!t]
\vspace{.2cm}
\includegraphics[width=8.1cm]{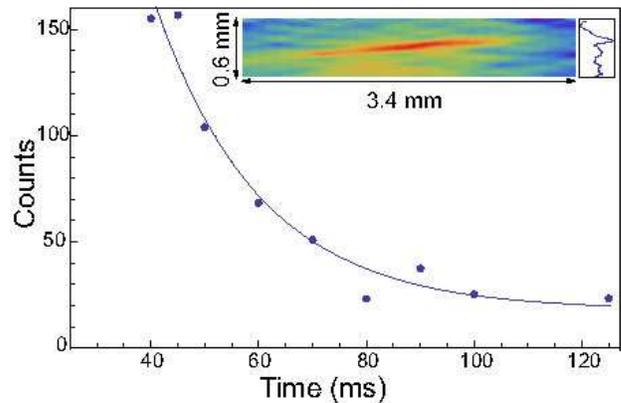}
\vspace{-.5cm}
\caption{(color online) Measurement of trap decay versus time. The curve is a least squares fit to the function  $b+ae^{-t/\tau}$ resulting in   $a= 1100, b=18,$ and 
$\tau=20~\rm ms$. The inset shows the spatial distribution at  $t=40~\rm ms$ 
formed by averaging 200, $1~\rm ms$ exposures. To the right is a transverse line profile through the center of the trap.}
\label{fig.trapdecay}
\end{figure}

We then used the BoB to trap Cs atoms from a MOT. The atoms in the MOT had an initial  temperature of about 50 $\mu$K measured using a time of flight method.  With our power and beam sizes, this gives the parameters shown in Table 1. 
To look for trapping we allowed the MOT to load for 1 s then turned on the bottle beam allowing both to overlap for about 1 ms. We then turned off the MOT beams for a  time $t$ allowing the MOT to fall away. Finally the MOT beams were briefly turned on to take a flourescence image of the atoms using a cooled EMCCD camera.  By varying the delay time $t$ between extinction of the MOT beams and measuring the atom number we extracted an exponential 
trap lifetime of $\tau\sim 20~\rm ms$, as shown in Fig. \ref{fig.trapdecay}. 
The atom distribution in the inset is consistent with a pencil of atoms of size 
$2 z_m\times 2\rho_m.$ This is expected since at $t=40~\rm ms$
 the atoms have been heated above the saddle temperature and essentially fill the volume between the axial and radial peaks.

The finite trap lifetime is due to several factors\cite{Savard1997} including  heating due to photon scattering, spatial noise of the trapping beams, and collisions with background atoms. 
We believe the dominant factor limiting the lifetime is  
substantial initial heating of the atoms by the non-adiabatic transfer from the MOT to the BoB trap
followed by photon scattering.
An initial  temperature of $500~\mu\rm K$ would imply a mean scattering rate 
of $5\times 10^3~\rm s^{-1}$ which in turn would heat the atoms 
to 1 mK in a time of 20 ms. Since the escape barrier at the saddle is $\sim 1~\rm  mK$, this estimate is consistent with the observed lifetime of $\tau=20~\rm ms.$

In conclusion we have demonstrated trapping of cold Cs atoms in a novel  BoB trap.
Our method only requires access through a single window, is relatively simple, and uses   standard optical components. It has the potential for sub $\mu\rm m$ trapping in 3D with very modest optical power.  Atoms were visible in our trap for more than 100 ms, and we believe that with a larger detuning and the addition of a cooling phase after trap loading,  lifetimes  limited by collisions with background atoms should be possible. In future work we will explore methods for cooling inside the BoB trap, as in Ref. \cite{Winoto1999}. Finally we note that this technique can be easily generalized to create other types of dark traps. 
By putting $P_2<P_1 w_2^2/w_1^2$ and $q>1$ there will be finite intensity at the origin, but an intensity null at $z=0,$ 
 $\rho=w_1 \frac{q}{q^2-1}\sqrt{\ln(q\sqrt{P_1/P_2})}.$ 
In this way microscopic toroidal traps can be generated which may be useful for studies of persistent currents and superfluid flow with cold atoms\cite{Ryu2007}. 

 This work was supported by the NSF and ARO-IARPA.

\pagebreak 

\end{document}